\documentclass[pre,superscriptaddress,showpacs,twocolumn]{revtex4}

\usepackage{graphicx}
\usepackage{amsmath}
\usepackage{amsfonts}
\usepackage{amssymb}

\usepackage{color}

\begin{document}

\title{Transient ageing in fractional Brownian and Langevin equation motion}

\author{Jochen Kursawe}
\affiliation{Wolfson Centre for Mathematical Biology, Mathematical Institute,
University of Oxford, 24-29 St Giles', Oxford OX1 3LB, United Kingdom}
\author{Johannes Schulz}
\affiliation{Physics Department, Technical University of Munich, 85747 Garching,
Germany}
\author{Ralf Metzler}
\affiliation{Institute of Physics \& Astronomy, University of Potsdam, 14776
Potsdam-Golm, Germany}
\affiliation{Department of Physics, Tampere University of Technology, FI-33101
Tampere, Finland}
\affiliation{Wolfson Centre for Mathematical Biology, Mathematical Institute,
University of Oxford, 24-29 St Giles', Oxford OX1 3LB, United Kingdom}

\begin{abstract}
Stochastic processes driven by stationary fractional Gaussian noise, that is,
fractional Brownian motion and fractional Langevin equation motion, are usually
considered to be ergodic in the sense that, after an algebraic relaxation,
time and ensemble averages of physical observables coincide. Recently it was
demonstrated that fractional Brownian motion and fractional Langevin equation
motion under external confinement are transiently non-ergodic---time and
ensemble averages behave differently---from the moment when the particle starts
to sense the confinement. Here we show that these processes also exhibit
transient ageing, that is, physical observables such as the time averaged mean
squared displacement depend on the time lag between the
initiation of the system at time $t=0$ and the start of the measurement at the
ageing time $t_a$. In particular, it turns out that for fractional Langevin
equation motion the ageing dependence on $t_a$ is different between the cases
of free and confined motion. We obtain explicit analytical expressions for the
aged moments of the particle position as well as the time
averaged mean squared displacement and present a numerical
analysis of this transient ageing phenomenon.
\end{abstract}


\maketitle

\section{Introduction}

Normal Brownian diffusion processes are stationary in the sense that physical
quantities such as two point correlation functions $C(t_1,t_2)$ only depend on
the difference of the two times $t_1$ and $t_2$, $C(t_1,t_2)=C(|t_1-t_2|)$.
Moreover, they are independent of the time difference $t_a$ between initiation
of the process at $t=0$ and start of the measurement at the \emph{ageing time\/}
$t_a$. For anomalous diffusion processes, characterized by a mean squared
displacement (MSD) of the form $\langle x^2(t)\rangle\simeq K_{\alpha}t^{\alpha}$
with the generalized diffusion coefficient $K_{\alpha}$ of physical dimension
$\mathrm{cm}^2/\mathrm{sec}^{\alpha}$, stationarity is not necessarily
fulfilled. In both the subdiffusive ($0<\alpha<1$) and superdiffusive
($\alpha>1$) regimes, physical observables may be non-ergodic and exhibit an
explicit dependence on the ageing time $t_a$.

Such consequences of non-stationary behavior may be illustrated by means of
the time averaged MSD
\begin{equation}
\label{eq:tamsd}
\overline{\delta^2(\Delta)}=\frac{1}{T-\Delta}\int\limits_{t_a}^{t_a+T-\Delta}
\Big[x(t+\Delta)-x(t)\Big]^2 dt.
\end{equation}
It evaluates a measured single particle trajectory in terms of the sliding
average along the time series $x(t)$ of length $T$ over the squared position
differences as separated by the lag time $\Delta$. This definition of the time
averaged MSD is typically used to analyze trajectories recorded in experiments
or simulations \cite{pt,igor}. Here we denote time averages by an
overline, $\overline{\,\,\,\cdot\,\,\,}$.  For trajectories of finite length,
the additional average over sufficiently many trajectories with label $i$,
$\langle\overline{\delta^2(\Delta)}\rangle=( 1/N)\sum_{i=1}^N
\overline{\delta^2(\Delta)}_i$, provides a smooth functional behavior.
The quantity $\overline{\delta^2( \Delta)}$ contrasts the more standard
ensemble MSD $\langle x^2(t)\rangle=\int x^2P(x,t)dx$, defined as the
spatial average of $x^2$ over the probability density function $P(x,t)$.

The integrand in expression (\ref{eq:tamsd}) for the time averaged MSD $\langle
\overline{\delta^2(\Delta)}\rangle$ of a random walk process may be
expressed as the product of the typical
squared length per jump, $\langle \delta x^2\rangle$, and the average number
of jumps $\langle n(t+\Delta,t)\rangle$ performed in the time interval between
$t$ and $t+\Delta$ \cite{he,johannes}. For Brownian motion, the number of
jumps is given by the length of the interval, $\Delta$, divided by the typical
time $\tau$ per jump, that is, $\langle n(t+\Delta,t)\rangle=\Delta/\tau$.
Defining the diffusion constant as $K_1=\langle\delta x^2\rangle/[2\tau]$,
after integration over $t$ we find that $\langle\overline{\delta x^2(\Delta)}
\rangle=2K_1\Delta$ for any finite $T$ \cite{pccp}. This result is independent
of the ageing time $t_a$ and the length $T$ of the time series: we say that the
process does not age. Moreover, it is ergodic in the sense that $\langle\delta
^2(\Delta)\rangle=\langle x^2(\Delta)\rangle$. For sufficiently large $T$ the
process is self-averaging in the sense that on average each jump occurs after
the time increment $\tau$, and thus a single trajectory $x(t)$ produces the 
result $\overline{\delta x^2(\Delta)}=2K_1\Delta$ such that the time averaged
MSD of different trajectories is fully reproducible \cite{he,pccp}.

In contrast, processes described by the Scher-Montroll continuous time random
walk (CTRW) model with a power-law distribution $\psi(\tau)\simeq\tau^{-1-
\alpha}$ ($0<\alpha<1$) of waiting times $\tau$ between successive jumps
\cite{scher} exhibit a so-called weak ergodicity breaking \cite{web}, $\langle
\delta^2(\Delta)\rangle\neq\langle x^2(\Delta)\rangle$: while the ensemble
averaged MSD scales sub-linearly with time, $\langle x^2(t)\rangle\simeq K_{
\alpha}t^{\alpha}$, the time averaged MSD is linear in the lag time, $\langle
\overline{\delta^2(\Delta)}\rangle\simeq K_{\alpha}\Delta/T^{1-\alpha}$ but
depends on the measurement time $T$ \cite{he,ariel}. Under confinement,
$\langle x^2(t)\rangle$ reaches the thermal plateau value $\langle x^2\rangle_{
\mathrm{th}}$ \cite{thermal}, while $\overline{\delta^2(\Delta)}\simeq(\Delta/
T)^{1-\alpha}$ continues to grow in power-law fashion
\cite{pnas,neusius}. Such weakly non-ergodic dynamics in absence and presence of
confinement was indeed observed experimentally in both the walls and the bulk
volume of living biological cells \cite{weigel,lene,stas}. Interestingly, a
similar weak
ergodicity breaking is obtained for time- and space-correlated CTRWs \cite{corr}
and `superageing' systems \cite{michael_bus} as well as for Markovian diffusion
with scaling forms of the position dependence of the diffusion coefficient
\cite{andrey}.  In superdiffusive systems the ergodic violation becomes
`ultraweak' in the sense that $\langle x^2(\Delta)\rangle$ and
$\langle\overline{\delta^2(\Delta)} \rangle$ differ only by a constant factor
\cite{zukla}. Moreover, the ageing dependence on $t_a$ was derived
for subdiffusive CTRW, in which the ageing time $t_a$ explicitly enters the
expressions for both $\langle x^2(t)\rangle$ and $\overline{\delta^2(\Delta)}$
\cite{eli_age,johannes}.  For the time averaged MSD the ageing time
$t_a$ enters solely through a multiplicative factor, independent of the process
details \cite{johannes}.

Here we are interested in the ageing properties of another popular class of
anomalous diffusion models:
fractional Brownian motion (FBM) and fractional Langevin equation (FLE) motion
are driven by stationary Gaussian but power-law correlated non-white noise and
algebraically reach ergodic behavior, $\overline{\delta^2(\Delta)}\sim\langle
x^2(\Delta)\rangle$ in the case of free, unconfined motion \cite{deng} and
$\overline{\delta^2(\Delta)}\sim2\langle x^2\rangle_{\mathrm{th}}$ for
confined motion \cite{jae,jeon1,REM}. FBM and FLE motion in the overdamped
regime behave similarly, and both emanate as
effective one-particle description in many-particle systems such as
viscoelastic environments, single file diffusion, or the motion of a monomer in
a long polymer chain \cite{hoefling,goychuk,szym,tobias}.
The ergodic nature of FBM/FLE anomalous diffusion was
observed for the motion of telomeres in cell nuclei \cite{bronstein}, the
diffusion of tracers in the cytoplasm of biological cells at sufficiently long
times \cite{weber,szym,lene}, or for the motion of simulated lipid molecules in
bilayers in different physical phases \cite{jae_lipid}. However, in contrast to
the exponential relaxation $\langle x^2(t)\rangle\sim\langle x^2\rangle_{\mathrm{
th}}(1-\mathrm{const.}t^{\alpha-2}\exp\{-\mathrm{const.'}t\})$ of the ensemble
averaged MSD under confinement, the time averaged MSD of FBM 
exhibit the power-law approach $\overline{\delta^2(\Delta)}\sim2\langle
x^2\rangle_{\mathrm{th}}(1-\mathrm{const.}\Delta^{\alpha-2})$ \cite{jeon1}. We call this a 
transiently non-ergodic behavior. Such a power-law relaxation of the time
averaged MSD was recently observed experimentally for microbeads tracked by
optical tweezers in a wormlike micellar solution \cite{lene1}.

Similar to the observation of weakly non-ergodic behavior in FBM or FLE motion
also the transient ageing behavior of these processes may become relevant under
certain experimental conditions, such that the influence of the ageing time
$t_a$ remains detectable during the measurement. Given the temporal resolution
of modern spectroscopic or single particle tracking methods, this may be
possible. Even more so, this question may become relevant in simulations studies
in which extremely high time resolution is possible. Here we derive the
different patterns of transient ageing in the FBM/FLE family of processes fueled
by power-law correlated Gaussian noise. Thus we find
that the time averaged MSD splits into two additive terms: a stationary term
depending solely on the lag time $\Delta$, and an ageing term that decays with
both ageing time $t_a$ and process time $T$. For both FBM and FLE motion the
dependence on the process time $T$ is inversely proportional. For confined FBM
we obtain an exponentially fast decay in $t_a$, while for FLE motion we find an
algebraic decay in the ageing time. Interestingly, the scaling exponent is
different between free and confined FLE motion.

The paper is structured as follows. After a definition of fractional Gaussian
noise and an introduction to the FLE and FBM in terms of their dynamic equations
in Sec.~\ref{sec_proc}, in Secs.~\ref{ffle} to \ref{fbm} we discuss the ensemble
and time averaged moments as well as the ergodic and ageing behavior of free
and confined FLE motion and confined FBM. In Sec.~\ref{disc} we draw
our conclusions. In the Appendices we describe the simulations procedure to
create stochastic trajectories $x(t)$ and collect results for the
autocorrelation functions.

\section{Anomalous diffusion models with power-law correlated, Gaussian noise}
\label{sec_proc}

We investigate two variants of stochastic differential equations, both driven
by so-called fractional Gaussian noise. These comprise the FLE,
in which the fluctuation-dissipation relation \cite{kubo} is satisfied, and
FBM, in which the noise is external and the motion therefore is not thermalized.
The associated
simulations schemes are compiled in Appendix \ref{sec:simulation}.

Fractional Gaussian noise, the derivative process of
FBM \cite{mandelbrot,gripenberg}, has zero mean $\langle\xi_H(t)\rangle=0$,
and the variance ($0<H<1$)
\begin{eqnarray}
\nonumber
\langle\xi_H(t_1)\xi_H(t_2)\rangle&=&2D_HH(2H-1)|t_1-t_2|^{2H-2}\\
&+&4D_HH|t_2-t_1|^{2H-1}\delta(t_2-t_1),
\label{eq:FGN}
\end{eqnarray}
exhibiting a power-law decay with scaling exponent $2H-2$ of the difference
between the two times $t_1$ and $t_2$. Here, $H$ denotes the Hurst exponent.
Integration of fractional Gaussian noise over time results in FBM with the anomalous diffusion exponent
$\alpha = 2H$. Here, we keep the notation in terms of $H$ for traditional
reasons. The $\delta$ term in Eq.~(\ref{eq:FGN}) guarantees that the
correlations converge in the limit $t_1=t_2$. Above functional form $\langle
\xi_H(t_1)\xi_H(t_2)\rangle=f(|t_1-t_2|)$ demonstrates that fractional Gaussian
noise is a stationary process. When $2H<1$ the noise correlator has a negative
sign (antipersistence), while positive correlations (persistence) occur for
$2H>1$. Normal Brownian noise without correlations corresponds to the limit
$H=1/2$, and $H=1$ is the limiting case of ballistic
(fully persistent) motion.

\subsection{Fractional Langevin equation (FLE)}

FLEs \cite{eric} are based on the generalized
Langevin equation with power-law memory kernel. Generalized Langevin equations
were particularly promoted by the work of Kubo \cite{kubo}, and have become a
standard tool to describe complex diffusion dynamics. In particular, FLEs have
been applied to model the internal dynamics of single protein molecules
\cite{min,kou}, or to describe diffusion in a
viscoelastic continuum \cite{goychuk}. FLEs were analyzed in the framework of
heat bath models in Ref.~\cite{kupferman}. Without an external potential, the
free FLE for the time-dependence of the position coordinate $y(t)$ reads
\begin{equation}
\label{eq:ffle}
m\frac{d^2}{dt^2}y(t)=-\overline{\gamma}\int\limits_0^t(t-\tau)^{2H-2}\frac{
dy}{d\tau}(\tau)d\tau+\eta\xi_H(t),
\end{equation}
valid for persistent Hurst exponents $1/2<H<1$. Here, $m$ denotes the mass of
the diffusing particle, $\overline{\gamma}$ is a generalized friction constant,
$\eta$ is the noise strength, and $\xi_H$ is the fractional Gaussian noise.
$\eta$ and $\overline{\gamma}$ are related by the fluctuation dissipation
theorem
\cite{kubo}
\begin{equation}
\eta=\sqrt{\frac{k_B\mathcal{T}\overline{\gamma}}{2D_HH(2H-1)}},
\end{equation}
where $k_B\mathcal{T}$ denotes thermal energy. Note that the power-law tail of
the fractional Gaussian noise (\ref{eq:FGN}) is matched by the power-law kernel
inside the memory integral expression of the FLE (\ref{eq:ffle}). For this
power-law form we may rewrite the FLE as
\begin{equation}
m\frac{d^2}{dt^2}y(t)=-\overline{\gamma}\Gamma(2H-1)\,_0^CD_t^{2-2H}y(t)+\eta
\xi(t),
\end{equation}
where the fractional operator $_0^CD_t^{2-2H}$ in the Caputo sense is defined
by comparison with Eq.~(\ref{eq:ffle}). Introducing dimensionless quantities
one can show that the only free parameter of Eq.~(\ref{eq:ffle}) is the initial
velocity $v_0$ in units of the thermal velocity $\sqrt{k_B\mathcal{T}/m}$, as
can also be seen from the full solution \eqref{eq:sec_mom_FFLE}.

In the presence of an external harmonic potential $V(x)=\frac{1}{2}\lambda x^2$,
the FLE assumes the form
\begin{equation}
\label{eq:flelambda}
m\frac{d^2}{dt^2}y(t)=-\overline{\gamma}\int\limits_{0}^t(t-\tau)^{2H-2}
\frac{dy}{d\tau}d\tau+\eta\xi_H(t)-\lambda y(t).
\end{equation}
with the Hookean restoring force term $\lambda y(t)$.
Neglecting the left hand side of this equation, we reach the overdamped limit.
Kou applied this overdamped FLE to the movements of ligands within a
single protein molecule \cite{kou}, and provided the equilibrium solution to
this equation.

\subsection{Fractional Brownian motion}

An alternative combination of fractional Gaussian noise and the particle
position $x$ is provided by the stochastic differential equation
\begin{equation}
\label{fbm_le}
\frac{dx(t)}{dt}=\xi_H(t).
\end{equation}
This FBM was introduced by Kolmogorov \cite{kolmogorov} and became famous
following the work of Mandelbrot and van Ness \cite{mandelbrot}. Further
analysis in terms of the stochastic formulation and the ergodicity of the
process are found in Refs.~\cite{deng,jae,jeon1}.  In the FBM model, the
potential is directly added to the derivative process of FBM. In case of a
Hookean force of strength $\lambda$ the stochastic equation becomes
\begin{equation}
 \label{eq:fbmk}
 \frac{dx(t)}{dt}=\xi_H(t)-\lambda x(t).
\end{equation}
Note that despite the seemingly simple structure of
Eqs.~(\ref{fbm_le}) and (\ref{eq:fbmk}) FBM is a strongly non-Markovian process
and cannot be mapped onto a random walk process \cite{weron}.

\begin{figure*}
\includegraphics{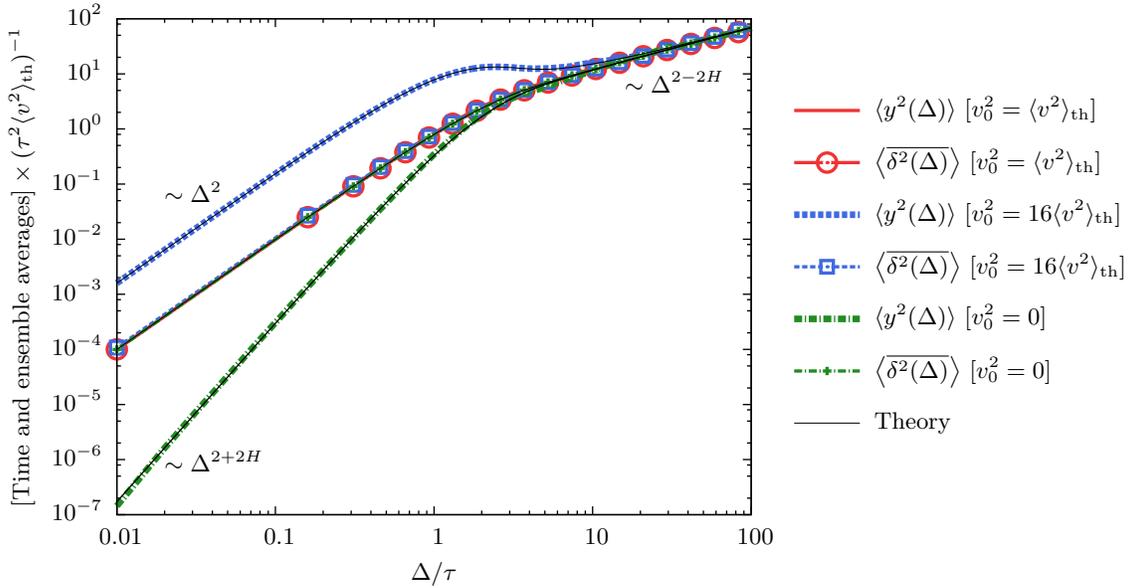}
\caption{Ensemble and time averaged MSDs for free FLE motion with different
initial velocities $v_0$. The solid lines represent the theoretical
results, Eqs.~\eqref{eq:freeflemsd} and \eqref{eq:FFLEf_st}. The
equilibrium MSD and the time averaged MSDs coincide. The ensemble
averages for the non-equilibrium processes approach the time averages at
long times $\Delta$. All simulations were performed with $h=0.01$ (See
Appendix \ref{sec:simulation}), $H=0.625$, $t_a=0$, and $T/\tau=100$.
Ensemble averages contain 1,000 trajectories.}
\label{fig:freeflemsd}
\end{figure*}

\section{Free fractional Langevin equation motion and ageing behavior}
\label{ffle}

In this and the following Sections we present the solutions for the ensemble
and time averaged MSDs of the FLE and FBM processes introduced here and investigate their
ageing behavior. Solutions to the FLE share many general properties with FBM
on longer time scales but due to their formulation in phase (velocity and
position) space have differentiable trajectories \cite{deng,pottier}.
We find that the ageing of FBM and FLE is significantly different. Remarkably,
the ageing of the FLE varies between the cases of free and confined motion.

\subsection{Ensemble averaged moments}

The free FLE has been widely used in literature, see for example Refs.
\cite{eric,hoefling,goychuk}, and an equilibrium solution was provided by
Pottier \cite{pottier}. Here, we extend this solution to non-equilibrium
initial conditions $v_0$ analogous to Deng and Barkai \cite{deng}. From the
two-point correlator, corresponding to the fluctuation [$\xi(t)$] average of
Eq.~(\ref{eq:ffle}) shown in Appendix \ref{app_ex} we obtain the second moment
\begin{eqnarray}
\nonumber
\langle y^2(t)\rangle&=&2\frac{k_B\mathcal{T}}{m}t^2E_{2H,3}(-\gamma t^{2H})\\
&&+\left(v_0^2-\frac{k_B\mathcal{T}}{m}\right)t^2E^2_{2H,2}(-\gamma t^{2H})
\label{eq:freeflemsd}
\end{eqnarray}
with $\gamma=\overline{\gamma}\Gamma(2H-1)/m$. The generalized Mittag-Leffler
function $E_{\alpha,\beta}$ has the following series expansions \cite{bateman}
\begin{subequations}
\begin{equation}
E_{\kappa,\rho}(z)=\sum_{n=0}^{\infty}\frac{z^n}{\Gamma(\rho+\kappa n)}
\end{equation}
and
\begin{equation}
E_{\kappa,\rho}(z)=-\sum_{n=1}^{\infty}\frac{z^{-n}}{\Gamma(\rho-\kappa n)}
\end{equation}
\end{subequations}
around $z=0$ and $|z|=\infty$, respectively. The free FLE
process thus exhibits a characteristic turnover from
ballistic motion
\begin{equation}
\label{ffle_init}
\langle y^2(t)\rangle\sim v_0^2t^2
\end{equation}
to subdiffusion (note that $H>1/2$),
\begin{equation}
\label{ffle_as}
\langle y^2(t)\rangle\sim\frac{2}{\gamma}\langle v^2\rangle_{\mathrm{th}}
\frac{t^{2-2H}}{\Gamma(3-2H)},
\end{equation}
where the thermal squared velocity is given by $\langle v^2\rangle_{\mathrm{th}}
=k_B\mathcal{T}/m$. With the anomalous diffusion constant $D_H=\langle v^2
\rangle_{\mathrm{th}}/[\gamma\Gamma(3-2H)]$ we rewrite the latter result in the
more convenient form
\begin{equation}
\label{eq:FFLE_long_times}
\langle y^2(t)\rangle\sim2D_Ht^{2-2H}.
\end{equation}
This result demonstrates that for FLE motion persistent (positively correlated)
fractional Gaussian noise effects subdiffusion, in contrast to the case of FBM
shown below. Result (\ref{eq:freeflemsd}) and the asymptotic behaviors
(\ref{ffle_init}) and (\ref{ffle_as}) are shown in Fig.~\ref{fig:freeflemsd},
with excellent agreement between theory and stochastic simulations. The
turnover between initial ballistic motion and terminal subdiffusion defines the
intrinsic time scale $\tau=\gamma^{-1/(2H)}$. 
Note that in the case of vanishing initial velocity the
ballistic motion is replaced by an initial hyperdiffusive regime
\begin{equation}
\label{fle_msd}
\langle y^2(t)\rangle\sim2(2H+1)\gamma\langle v^2\rangle_{\mathrm{th}}\frac{
t^{2+2H}}{\Gamma(3+2H)}, 
\end{equation} 
which is also shown in Fig.~\ref{fig:freeflemsd}. The possibility of such a
superballistic diffusion for Langevin equation models with vanishing initial
velocities was reported in Ref.~\cite{siegle}, showing that superballistic
diffusion may occur due to the high amount of energy that the particle gains
during equilibration.  Notably, this effect occurs even for a Brownian particle
($H=1/2$).

According to Eq.~(\ref{eq:freeflemsd}), the mean squared position of the free
FLE process has a non-equilibrium contribution that is proportional to the
difference $v_0^2-\langle v^2\rangle_{\mathrm{th}}$ between initial and thermal
kinetic energy. This contribution vanishes algebraically at long times,
mirroring the equilibration but also the ageing of the process, see below.

The first moment of $y$,
\begin{equation}
\langle y(t)\rangle=v_0tE_{2H,2}(-\gamma t^{2H}),
\label{eq:first_moment_FFLE}
\end{equation}
has the asymptotic behavior
\begin{equation}
\langle y(t)\rangle\sim v_0t
\end{equation}
at short times and, terminally,
\begin{equation}
\langle y(t)\rangle\sim\frac{v_0}{\gamma\Gamma(2-2H)}t^{1-2H}.
\end{equation}
It thus approaches zero at long times since the equilibrated process is
symmetric. Intriguingly, the approach to the steady state becomes slower when
the Hurst parameter tends to 1/2.
The Brownian case $H=1/2$ itself equilibrates exponentially, and at
long times has the thermal first moment of $\lim_{t\to\infty}\langle
y(t)\rangle=v_0/\gamma$. This property can be interpreted as follows: the
Brownian particle starting with an initial velocity will keep this velocity for
a while until the friction slows it down on the characteristic length scale
$v_0/\gamma$. Afterwards, the particle diffuses symmetrically
around this stalling point. A free FLE particle with its coupled power-law forms
of the noise correlations and the friction kernel does not feature this
property and will on long time scales diffuse around the original starting
point.

Combining the second and first moments we obtain the variance $\langle y^2(t)
\rangle-\langle y(t)\rangle^2$. At short times, this quantity is independent of
the initial velocity $v_0$ and exhibits precisely the hyperdiffusive behavior
(\ref{fle_msd}). At longer times, due to the thermalization of the velocity,
the asymptotic behavior of $\langle[\Delta y(t)]^2\rangle$ follows expression
(\ref{eq:FFLE_long_times}).

\begin{figure}
\includegraphics{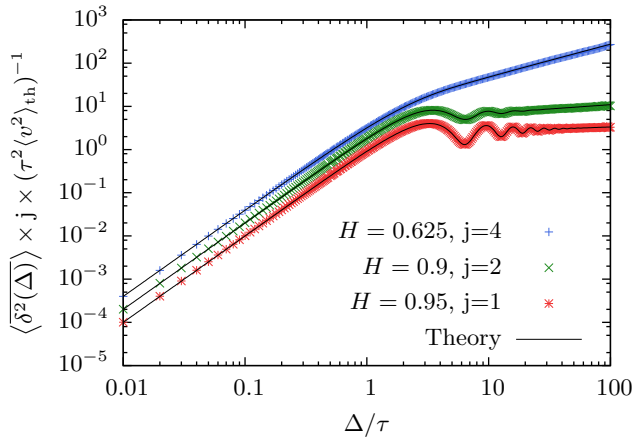}
\caption{For large Hurst parameters, $H>0.8$,
oscillations occur in the MSD of the free FLE. These oscillations are more
pronounced for larger $H$. The data was simulated for different Hurst
parameters with initial velocities $v_0^2= \langle v^2\rangle_{\mathrm{th}}$,
$h=0.01$ (see Appendix \ref{sec:simulation}), $T/\tau=100$, and for 1,000
trajectories. The theoretical values are calculated according to
Eq.~\eqref{eq:FFLEf_st}. For better visibility, the plots are shifted by the
factor j along the logarithmic ordinate.}
\label{fig:oscillations}
\end{figure}

An additional feature of the FLE process are distinct oscillations in the MSD,
as shown in Fig.~\ref{fig:oscillations}. These were reported before
\cite{jeon1,elistas}
and appear for large Hurst index, and they occur for all possible initial
conditions in both the ensemble and time averaged MSD. For which values of the
Hurst parameter $H$ are these oscillations possible? The derivative of
Eq.~\eqref{eq:freeflemsd} shows that the equilibrium MSD of the fee FLE only has
extrema and thus oscillates if the function $E_{2H,2}(-\gamma t^{2H})$ has
zeros. Numerical analysis demonstrates that this condition is met for $H\gtrsim
0.8\pm 10^{-4}$. Remarkably, this critical Hurst parameter for oscillations is
preserved under non-equilibrium initial conditions.

\subsection{Time averaged mean squared displacement and ageing}

From the exact results for the second moment $\langle y^2(t)\rangle$ and the
autocorrelation function $\langle y(t_1)y(t_2)\rangle$ we obtain the time averaged MSD through
\begin{eqnarray}
\nonumber
\left<\overline{\delta^2(\Delta)}\right>&=&\frac{1}{T-\Delta}\int_{t_a}^{t_a+
T-\Delta}\Big(\langle x^2(t'+\Delta)\rangle+\langle x^2(t')\rangle\\
&&-2\langle x(t'+\Delta)x(t')\rangle\Big)dt'.
\label{EATAMSDnew}
\end{eqnarray}
Based on the full solution \eqref{eq:sec_mom_FFLE} for the free FLE, we find
that $\langle\overline{\delta^2(\Delta)}\rangle$ splits up into two additive
contributions,
\begin{equation}
\label{age_split}
\left<\overline{\delta^2(\Delta)}\right>=f_{\mathrm{st}}(\Delta)+f_{\mathrm{
age}}(\Delta;t_a,T).
\end{equation}
In contrast to
CTRW subdiffusion the time averaged MSD of the free FLE motion does not vanish
for long measurement and ageing times, $T$ and $t_a$. Namely, it has a
stationary contribution $f_{\mathrm{st}}$, that only depends on the lag time
$\Delta$. All dependencies on $t_a$ or $T$ are captured by the ageing term
$f_{\mathrm{age}}$, which decays for long $T$ and $t_a$. We will obtain an
analogous behavior for the FLE in an harmonic potential and FBM below.

\begin{figure}
\includegraphics{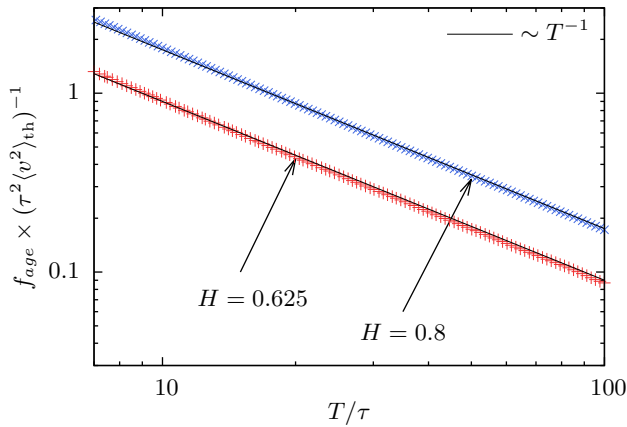}
\caption{Measurement time $T$ dependence of the free FLE, confirming the
proportionality to $1/T$. $f_{\mathrm{age}}$ was calculated by subtracting the
theoretical results for $f_{\mathrm{st}}$ \eqref{eq:FFLEf_st} from the simulated values for the time
averaged MSD. The simulations were run with $h=0.01$ (see
Appendix \ref{sec:simulation}), $v_0^2=100\langle v^2\rangle_{\mathrm{th}}$,
$\Delta/\tau=0.5$, and $t_a=0$, and were averaged over 1,000 trajectories.}
\label{fig:freefleagingT}
\end{figure}

The stationary contribution for the free FLE becomes
\begin{equation}
\label{eq:FFLEf_st}
f_{\mathrm{st}}=2\langle v^2\rangle_{\mathrm{th}}\Delta^2E_{2H,3}(-\gamma
\Delta^{2H}),
\end{equation}
with a turnover from the ballistic short time expansion
\begin{equation}
f_{\mathrm{st}}\sim\frac{2\langle v^2\rangle_{\mathrm{th}}}{\Gamma(2H)}\Delta^2
\end{equation}
to the asymptotic subdiffusive form
\begin{equation}
f_{\mathrm{st}}\sim\frac{2\langle v^2\rangle_{\mathrm{th}}}{\gamma\Gamma(2H-2)}
\Delta^{2-2H}.
\end{equation}
For the ageing term the following expression holds,
\begin{eqnarray}
\nonumber
f_{\mathrm{age}}&=&\frac{v_0^2-\langle v^2\rangle_{\mathrm{th}}}{T-\Delta}\int
\limits_{t_a}^{t_a+T-\Delta}\Big[(t+\Delta)E_{2H,2}(-\gamma(t+\Delta)^{2H})\\
&&\hspace*{3.2cm}-tE_{2H,2}(-\gamma t^{2H})\Big]^2dt,
\end{eqnarray}
which vanishes, when the initial velocity has the value of the thermal
velocity, $\pm\langle v^2\rangle_{\mathrm{th}}^{1/2}$. The term in the integral
is always positive and at long $t$ has the asymptotic form
\begin{equation}
\Big[\ldots\Big]^2\sim\frac{\Delta^2}{\gamma^2\Gamma^2(1-2H)}t^{-4H},
\end{equation}
such that due to $H>1/2$ it is integrable. We can further approximate the
ageing term in the experimentally relevant regime. For evaluating time averages
it is crucial to have sufficiently long time series, i.e. the collected data
should be extensive enough to produce reliable statistics.  That means, we will
observe the limit $T\gg\Delta$. Additionally, we assume the intrinsic time
scale to be very short in comparison to accessible measurement times, i.e. we
take the limit $T\gg\tau$. Finally, we keep the ageing time finite, i.e. $T\gg
t_a$. In these limits, the integral will approach a finite value,
$\daleth_{\infty}(\Delta,t_a)$, due to the above convergence criterion.  We
thus identify the long-$T$ dependence
\begin{equation}
f_{\mathrm{age}}=\left[v_0^2-\langle v^2\rangle_{\mathrm{th}}\right]\frac{
\daleth_{\infty}(\Delta,t_a)}{T}+\mathcal{O}\left(\gamma^{-2}T^{-4H}\right).
\end{equation}
The $1/T$-scaling of the ageing contribution $f_{\mathrm{age}}$ is demonstrated
in comparison with simulations results in Fig.~\ref{fig:freefleagingT}. For
$T\to\infty$ only the stationary term of $\langle\overline{\delta^2(\Delta)}
\rangle$ remains, and in that limit we find the ergodic behavior
\begin{eqnarray}
\nonumber
\left<\overline{\delta^2(\Delta)}\right>&=&\lim_{t\to\infty}\left<\left[y(t+\Delta)-y(t)\right]^2\right> \\
&=&\left< y^2(\Delta)
\right>,
\end{eqnarray}
under the condition that the initial velocity is thermalized, $v_0^2 = \langle v^2\rangle_{\mathrm{th}}$,
consistent with the results of Ref.~\cite{deng}.

\begin{figure}
\includegraphics{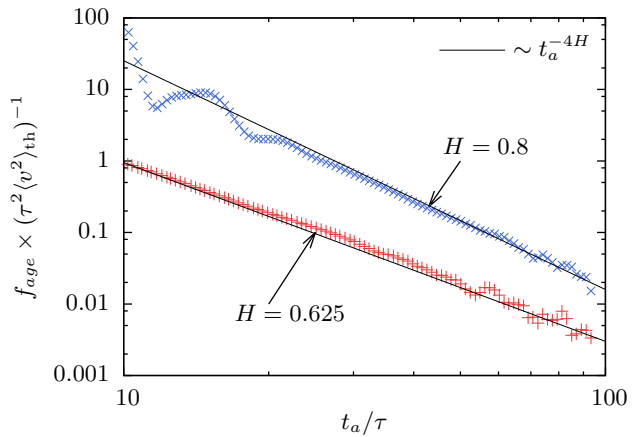}
\caption{Ageing time dependence of the free FLE motion, recovering the
proportionality to $t_a^{-4H}$. Parameters: $h=0.01$ (See Appendix
\ref{sec:simulation}), $\Delta/\tau=0.5$, and $T/\tau=5$, so that the data are
in the limit $t_a\gg T\gg\Delta$. For $H=0.625$, the initial velocity is $v_0^2
=4\times10^4\langle v^2\rangle_{\mathrm{th}}$ and the number of trajectories is
10,000.  For $H=0.8$, the initial velocity is $v_0^2=25\times10^4\langle v^2
\rangle_{\mathrm{th}}$ and the number of trajectories is $10^5$. The simulated
values for $H=0.8$ were multiplied by a constant factor of 10 for visual
convenience.}
\label{fig:freefleagingta}
\end{figure}

\begin{figure*}
\includegraphics{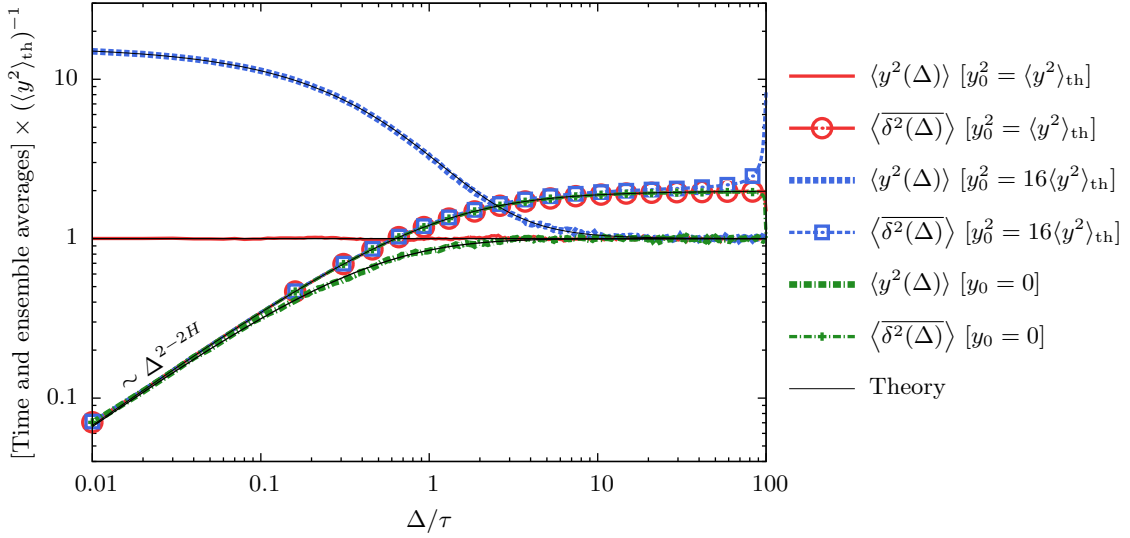}
\caption{Ensemble averaged and time averaged MSDs for the overdamped FLE motion
in a confining, external harmonic potential for $H=0.625$. The simulations were
performed with the parameter values $t_a=0$, $T/\tau=100$, and $h=0.01$ (see
Appendix \ref{sec:simulation}), and averaged over $10^4$ trajectories. The
theoretical behavior is provided by Eqs.~\eqref{eq:msdoflelambda} and
\eqref{eq:OFLElambdaf_st}. We observe excellent agreement between theory
and simulation. In accordance with Ref.~\cite{jeon1}, the time averages
approach twice the value of the ensemble averages.}
\label{fig:msdoflew}
\end{figure*}

What happens when the process is strongly aged? In the
associated limit of
$t_a\gg\Delta$ and $t_a\gg\gamma^{-1/(2H)}$ as well as $T\gg\Delta$ we obtain
\begin{equation}
f_{\mathrm{age}}=\frac{\left(v_0^2-\langle v^2\rangle_{\mathrm{th}}\right)
\Delta^2}{(1-4H)\gamma^2\Gamma^2(1-2H)}\frac{(T+t_a)^{1-4H}-t_a^{1-4H}}{T}.
\end{equation}
If additionally $t_a\gg T$, we thus find the scaling
\begin{equation}
\label{st_age}
f_{\mathrm{age}}\sim\frac{\left(v_0^2-\langle v^2\rangle_{\mathrm{th}}\right)
\Delta^2}{\gamma^2\Gamma^2(1-2H)}t_a^{-4H}.
\end{equation}
This asymptotic dependence on the ageing time $t_a$ is illustrated in
Fig.~\ref{fig:freefleagingta}.
Above results show that the relaxation of the ageing dependence becomes
slower when $H$ approaches the Brownian value $H=1/2$.

Remarkably, the first order correction for highly aged systems is quadratic in
the lag time $\Delta$. Hence, the coefficient for the ballistic regime in Eq.
\eqref{eq:FFLEf_st} could potentially be over- or underestimated if it is
gained from aged particle trajectories. 

\section{Confined fractional Langevin equation motion and ageing behavior}
\label{cfle}

We now turn to the FLE (\ref{eq:flelambda}) in the presence of a confining,
external harmonic potential and analyze the dynamics in terms of the ensemble
and time averaged MSDs. In particular, we find the remarkable result that
the ageing behavior is different from the free FLE motion.

\subsection{Ensemble averaged moments}

In the underdamped case [i.e., keeping the inertial term on the left hand
side of Eq. \eqref{eq:flelambda}], the second moment of the position can be
written in the form (compare Appendix \ref{app_ex})
\begin{eqnarray}
\nonumber
\langle y^2(t)\rangle&=&\langle y^2\rangle_{\mathrm{th}}+\left(y_0^2-
\langle y^2\rangle_{\mathrm{th}}\right)A^2(t)\\
&&\hspace*{-0.8cm}
+\left(v_0^2-\langle v^2\rangle_{\mathrm{th}}\right)B^2(t)+2y_0v_0A(t)B(t)
\label{dette}
\end{eqnarray}
where $\langle y^2\rangle_{\mathrm{th}}=k_B\mathcal{T}/\lambda$ is the thermal
value of the position coordinate, and $A(t)$ and $B(t)$ are defined via their
Laplace transforms ($s$ denoting the Laplace variable) through
\begin{subequations}
\begin{eqnarray}
A(s)&=\frac{\displaystyle\gamma s^{1-2H}+ms}{\displaystyle ms^2+\gamma s^{2-2H}
+\lambda}\\
B(s)&=\frac{\displaystyle m}{\displaystyle ms^2+\gamma s^{2-2H} +\lambda}.
\end{eqnarray}
\label{abbs}
\end{subequations}
In the overdamped case, which was discussed in Ref.~\cite{kou}, we obtain the
simpler result 
\begin{equation}
\label{eq:msdoflelambda}
\langle y^2(t)\rangle=\langle y^2\rangle_{\mathrm{th}}+\left(y_0^2-\langle y^2
\rangle_{\mathrm{th}}\right)E_{2-2H}^2\left(-\frac{\lambda}{\gamma}t^{2-2H}
\right)
\end{equation}
with $\gamma=\overline{\gamma}\Gamma(2H-1)$. Here $E_{\alpha}=E_{\alpha,1}$
is the regular Mittag-Leffler function. At short times we observe the leading
order behavior
\begin{equation}
\langle y^2(t)\rangle\sim y_0^2-\frac{2\lambda\left(y_0^2-\langle y^2\rangle_{
\mathrm{th}}\right)}{\gamma\Gamma(3-2H)}t^{2-2H}.
\end{equation}
That is, starting from the initial value $y_0^2$ we have a decay or increase of $\langle
y^2(t)\rangle$ depending on the sign of $y_0^2-\langle y^2\rangle_{\mathrm{th}}$.
Asymptotically the second moment exhibits the power-law convergence
\begin{equation}
\langle y^2(t)\rangle\sim\langle y^2\rangle_{\mathrm{th}}+\left(y_0^2-\langle
y^2\rangle_{\mathrm{th}}\right)\frac{\gamma^2}{\lambda^2}\frac{t^{4H-4}}{\Gamma
^2(2H-1)}
\end{equation}
to the thermal value $\langle y^2\rangle_{\mathrm{th}}=k_B\mathcal{T}/\lambda$.
The approach to this
thermal value is from above when $y_0^2>\langle y^2\rangle_{\mathrm{th}}$ and
from below in the opposite case. This relaxation occurs on the intrinsic time
scale $\tau=(\gamma/\lambda)^{1/(2-2H)}$. The above behavior is shown
in Fig.~\ref{fig:msdoflew}. We note that in terms of dimensionless variables
Eq.~\eqref{dette} has two free parameters: the rescaled initial
velocity and the initial position. In the overdamped case only the initial
position is relevant, see also the full solutions,
Eqs.~\eqref{eq:sec_mom_FLElambda} and \eqref{eq:sec_mom_ov_FLElambda}.

The first moment of the confined FLE motion is given by the relaxation pattern
\begin{equation}
\label{firstfle}
\langle y(t)\rangle=A(t)y_0+B(t)v_0
\end{equation}
for the full solution and, in the overdamped case, by the Mittag-Leffler
form
\begin{equation}
\label{eq:OFLE_lambda_first_moment}
\langle y(t)\rangle=y_0E_{2-2H}\left(-\frac{\lambda}{\gamma}t^{2-2H}\right)
\end{equation}
with the asymptotic behavior
\begin{equation}
\langle y(t)\rangle\sim\frac{y_0\gamma}{\gamma\Gamma(2H-1)}t^{2H-2}.
\end{equation}

\subsection{Time averaged mean squared displacement and ageing}

The full solution for the time averaged MSD $\langle\overline{\delta^2(\Delta)}
\rangle$ is additive as in Eq.~(\ref{age_split}) with the following two
contributions. The stationary term reads
\begin{equation}
f_{\mathrm{st}}=2\langle y^2\rangle_{\mathrm{th}}(1-A(\Delta))
\end{equation}
in the underdamped case and
\begin{equation}
\label{eq:OFLElambdaf_st}
f_{\mathrm{st}}=2\langle y^2\rangle_{\mathrm{th}}\left[1-E_{2-2H}\left(-\frac{
\lambda}{\gamma}\Delta^{2-2H}\right)\right]
\end{equation}
in the overdamped case. The latter turns over from the subdiffusive short lag
time growth
\begin{equation}
f_{\mathrm{st}}\sim2\langle y^2\rangle_{\mathrm{th}}\frac{\lambda}{\gamma}
\frac{\Delta^{2-2H}}{\Gamma(3-2H)}
\end{equation}
to the asymptotic long-$\Delta$ approach
\begin{equation}
f_{\mathrm{st}}\sim2\langle y^2\rangle_{\mathrm{th}}\left[1-\frac{\gamma}{
\lambda}\frac{\Delta^{2H-2}}{\Gamma(2H-1)}\right]
\end{equation}
to twice the thermal value $\langle y^2\rangle_{\mathrm{th}}$. This factor of
two stems from the very definition of the time averaged MSD (\ref{eq:tamsd}),
compare Eq.~(\ref{EATAMSDnew}).
\cite{jeon1}.

\begin{figure}
\includegraphics{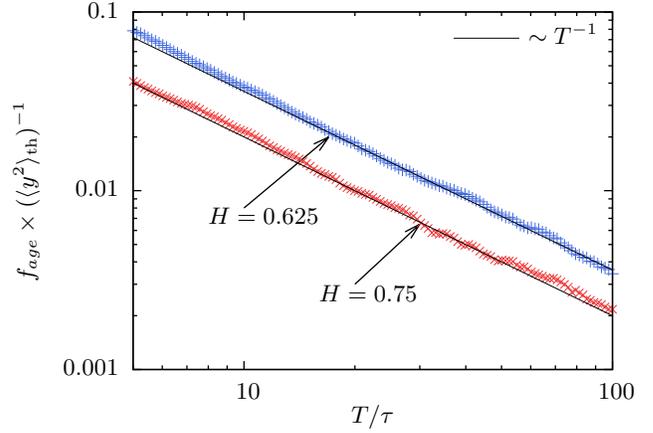}
\caption{Measurement time $T$ dependence of the
overdamped confined FLE process showing the predicted $T^{-1}$ decay.
$f_{\mathrm{age}}$ was obtained by subtracting the theoretical value of
$f_{\mathrm{st}}$ from the time averaged MSD. The simulation were performed
for $y_0^2=16\langle y^2\rangle_{\mathrm{th}}$, $h=0.01$ (see Appendix
\ref{sec:simulation}), $\Delta/\tau=0.3$, and $t_a=0$, and were averaged over
$10^4$ trajectories.}
\label{fig:agingToflew}
\end{figure}

The ageing term becomes
\begin{eqnarray}
\nonumber
f_{\mathrm{age}}&=&\frac{y_0^2-\langle y^2\rangle_{\mathrm{th}}}{T-\Delta}\int
\limits_{t_a}^{t_a+T-\Delta}\Big(A(t+\Delta)-A(t)\Big)^2dt\\
\nonumber
&&+\frac{v_0^2-\langle v^2\rangle_{\mathrm{th}}}{T-\Delta}\int\limits_{t_a}^{
t_a+T-\Delta}\Big(B(t+\Delta)-B(t)\Big)^2dt\\
\nonumber
&&+\frac{2y_0v_0}{T-\Delta}\int\limits_{t_a}^{t_a+T-\Delta}\Big(A(t+\Delta)-A(t)
\Big)\\
&&\hspace*{2.4cm}\times\Big(B(t+\Delta)-B(t)\Big)dt,
\end{eqnarray}
which can be used for numerical evaluation. In the overdamped case, this
expression simplifies to
\begin{eqnarray}
\nonumber
f_{\mathrm{age}}&=&\frac{y_0^2-\langle y^2\rangle_{\mathrm{th}}}{T-\Delta}
\int\limits_{t_a}^{t_a+T-\Delta}\left[E_{2-2H}\left(-\frac{\lambda}{\gamma}
(t+\Delta)^{2-2H}\right)\right.\\
&&\hspace*{2.4cm}\left.-E_{2-2H}\left(-\frac{\lambda}{\gamma}t^{2-2H}\right)
\right]^2dt
\end{eqnarray}
For $t\gg\Delta$ the integrand in the above expression decays like
\begin{eqnarray}
\nonumber
\Big[\ldots\Big]^2&\sim&\frac{\lambda^2}{\gamma^2}\Delta^2t^{2-4H}E_{2-2H,
2-2H}^2\left(-\frac{\lambda}{\gamma}t^{2-2H}\right)\\
&\sim&\frac{\gamma^2}{\lambda^2\Gamma^2(2-2H)}\Delta^2t^{4H-6},
\end{eqnarray}
where the second approximation holds when, in addition, $t\gg\tau$.
As the exponent $4H-6$ ranges in the interval between $-2$ and $-4$ the integral
converges and approaches the finite value $\daleth'_{\infty}(\Delta,t_a)$,
which increases with $t_a$. For $T\gg\Delta, \tau, t_a$ we thus find
\begin{equation}
f_{\mathrm{age}}\sim\left(y_0^2-\langle y^2\rangle_{\mathrm{th}}\right)\frac{
\daleth'_{\infty}(\Delta,t_a)}{T}.
\end{equation}
This $1/T$ dependence on the process time $T$ is in good agreement with our
simulations, as demonstrated in Fig.~\ref{fig:agingToflew}.

\begin{figure*}
\includegraphics{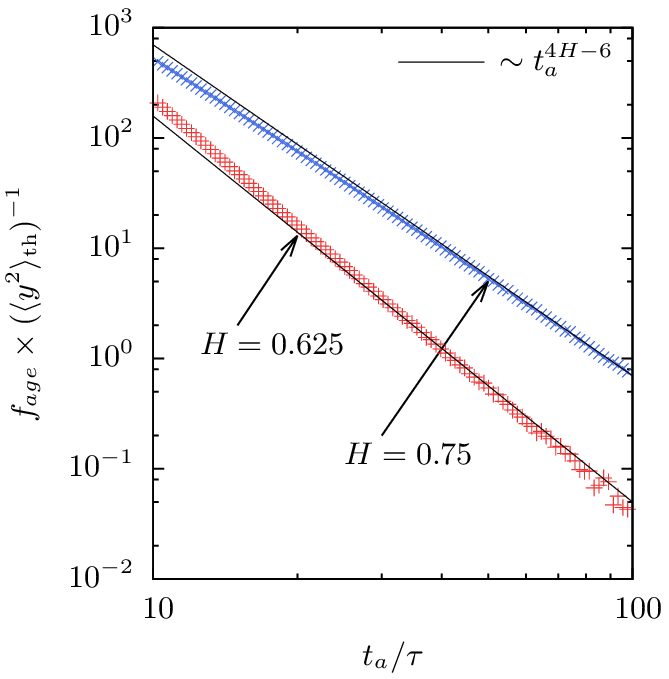}
\includegraphics{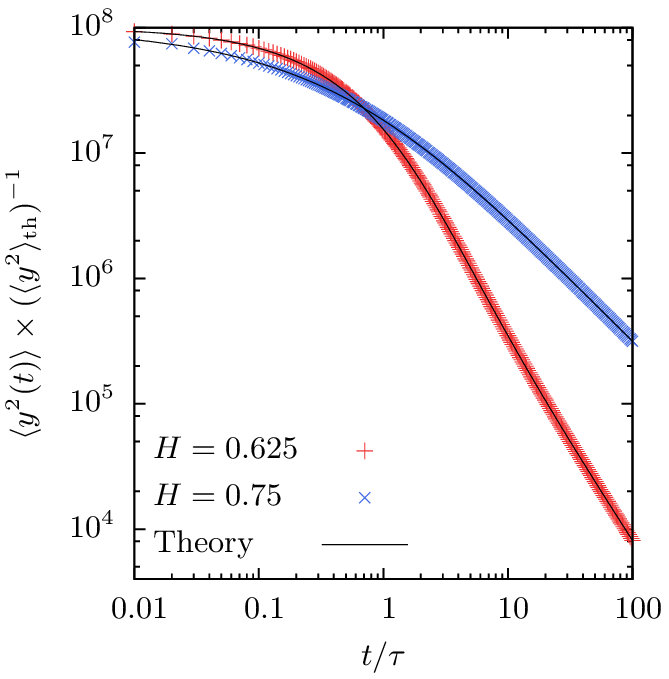}
\caption{Left: ageing dependence on $t_a$ of the overdamped
confined FLE process. The simulations used $y_0^2=10^8\langle
y^2\rangle_{\mathrm{th}}$ and $h=0.01$ (see Appendix \ref{sec:simulation}),
and the values are averaged over $10^4$ trajectories. We chose
$\Delta/\tau=0.3$ and $T/\tau=0.4$ to guarantee the limit $t_a\gg\Delta,T$.
In order to validate that the linearization in the simulation scheme is
accurate for this particular $h$ and the comparatively high starting kinetic
energies, the mean squared position is plotted on the right side and compared
to the theoretical result \eqref{eq:msdoflelambda}, observing excellent
agreement.}
\label{fig:agingtaoflewfigure}
\end{figure*}

In the ageing limit $t_a\gg\Delta$, $t_a\gg\tau$, and $T\gg\Delta$
we find the asymptotic behavior
\begin{equation}
f_{\mathrm{age}}\sim\frac{(y_0^2-\langle y^2\rangle_{\mathrm{th}})\gamma^2}{(4H
-5)\lambda^2\Gamma^2(2-2H)}\Delta^2\frac{(T+t_a)^{4H-5}-t_a^{4H-5}}{T}.
\end{equation}
If we assume strong ageing in the sense
$t_a\gg T$, we arrive at the scaling behavior
\begin{equation}
f_{\mathrm{age}}\sim\frac{(y_0^2-\langle y^2\rangle_{\mathrm{th}})\gamma^2}{
\lambda^2\Gamma^2(2-2H)}\Delta^2t_a^{4H-6}.
\end{equation}
This prediction is confirmed in
Fig.~\ref{fig:agingtaoflewfigure}. Note the difference to the
$t_a$-scaling (\ref{st_age}) of the free FLE motion. Remarkably, the ageing
terms for over- and underdamped FLE under confinement coincide in the
considered limits.

\section{Fractional Brownian motion}
\label{fbm}

We now turn to FBM, in which the fractional Gaussian noise is external.
Therefore, when we consider the ensemble averaged MSD of free FBM,
\begin{equation}
\label{fbm_free}
\langle x^2(t)\rangle=2D_Ht^{2H},
\end{equation}
we see that antipersistent noise with $0<H<1/2$ corresponds to subdiffusion,
in contrast to FLE motion, where the noise is counterweighted by the friction
kernel. Here, we calculate the dynamic quantities for confined FBM described by
Eq.~(\ref{eq:fbmk}). This derivation is analogous to Ref.  \cite{jeon1}. 

\subsection{Ensemble averaged moments}

\begin{figure*}
\includegraphics{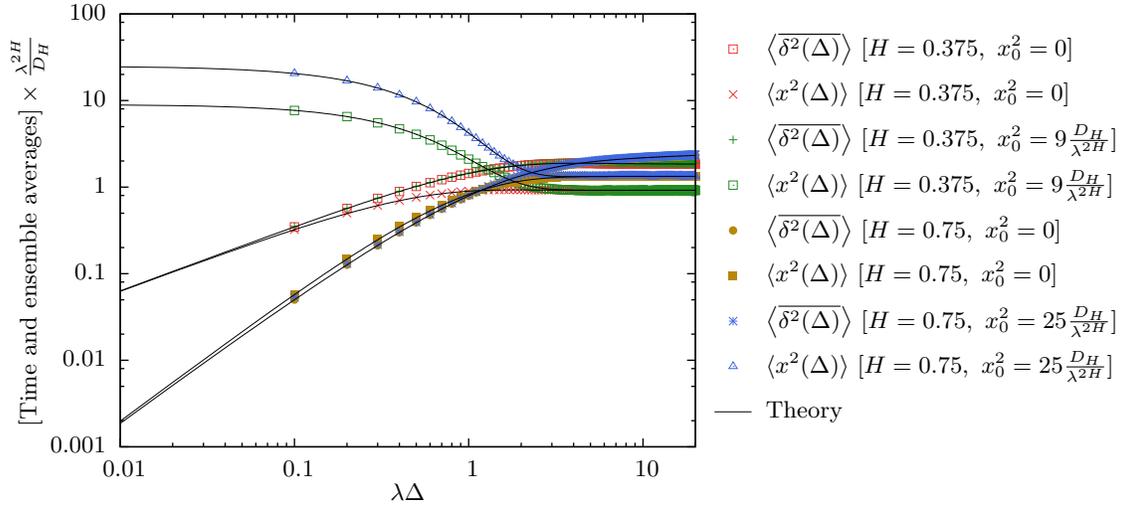}
\caption{Time and ensemble averaged MSDs, Eqs.~\eqref{eq:TAMSDstat} and
\eqref{eq:fbmkmsd} for confined FBM with the indicated parameters.  We observe
excellent agreement between theory and simulation. For a fixed Hurst parameter,
the time averaged MSD approaches twice the stationary value $\langle x^2\rangle
_{\mathrm{th}}$. The stationary values depend on the Hurst parameter
$H$, see Ref.~\cite{oleksii}. The short time behavior depends on the initial
value $x_0$. All simulations were run with $h=0.1$ (see Appendix
\ref{sec:simulation}), $T/\tau=300$, and $t_a=0$, and represent averages over
$10^5$ trajectories.}
\label{fig:msdfbmk}
\end{figure*}
 
As this process does not fulfill Kubo's fluctuation dissipation theorem, it
does not have a solution that starts in equilibrium. 
If one extends the solution of Ref.~\cite{jeon1} to
non-zero initial conditions one obtains the second moments listed in Appendix
\ref{app_ex}. The ensemble averaged MSD yields, 
\begin{eqnarray}
\nonumber
\langle x^2(t)\rangle&=&x_0^2e^{-2\lambda t}+\frac{D_H}{\lambda^{2H}}\gamma(
2H+1,\lambda t)+2D_He^{-\lambda t}t^{2H}\\
&&\hspace*{-0.8cm}-\frac{\lambda D_H}{2H+1}e^{-2\lambda t}t^{2H+1}M(2H+1,2H+2,
\lambda t). 
\label{eq:fbmkmsd}
\end{eqnarray}
Here, $\gamma$ is the lower incomplete $\gamma$-function, and $M$ denotes the
Kummer function \cite{abramowitz}. FBM in an harmonic external potential has an
intrinsic time scale of $\tau=1/\lambda$, see Fig.~\ref{fig:msdfbmk}.
Rescaling of the dynamic equation for the confined FBM reveals that the only
free parameter of this equation is the initial position.

At short times we obtain the behavior
\begin{equation}
\langle x^2(t)\rangle\sim x_0^2-2\lambda x_0^2t+2D_Ht^{2H}.
\end{equation}
For the case $H<1/2$ the
fractional term dominates, for $H>1/2$ the linear term dominates.
Counterintuitively, the process exhibits overshooting, i.e. local maxima in the
MSD, for the subdiffusive case $H<1/2$ (not shown).
In the limit $\lambda=0$ we recover
the ensemble averaged MSD (\ref{fbm_free}) of free FBM. With the asymptotic
expansions
\begin{equation}
M(2H+1,2H+2,z)\sim\frac{2H+1}{z}e^z
\end{equation}
and
\begin{equation}
\gamma(2H+1,z)\sim\Gamma(2H+1)-z^{2H}e^{-z}
\end{equation}
for the Kummer and incomplete $\gamma$-functions, respectively, in the limit
$t\gg \tau$ we find
\begin{eqnarray}
\langle x^2(t)\rangle&\sim&\frac{D_H\Gamma(2H+1)}{\lambda^{2H}}+x_0^2e^{-2
\lambda t}  \nonumber \\
&&-\frac{2}{\lambda^2}2H(2H-1)D_Ht^{2H-2}e^{-\lambda t}.
\end{eqnarray}

Thus, after an exponential
relaxation with characteristic decay time $1/\lambda$ the system reaches the
stationary value
\begin{equation}
\langle x^2\rangle_{\mathrm{st}}=D_H\Gamma(2H+1)/\lambda^{2H}.
\end{equation}
As the Kubo condition is missing for this process, the stationary value
  explicitly depends on the Hurst parameter $H$, thus demonstrating the
non-equilibrium character. The activation as function of $H$ of the stationary
value $\langle x^2\rangle_{\mathrm{st}}$ was demonstrated by
simulations in Ref.~\cite{oleksii}. For the first moment we find the
exponential decay 
\begin{equation}
\langle x(t)\rangle=x_0e^{-\lambda t}.
\label{eq:first_moment_fbmk}
\end{equation}

\subsection{Time averaged mean squared displacement and ageing}

\begin{figure}
\includegraphics{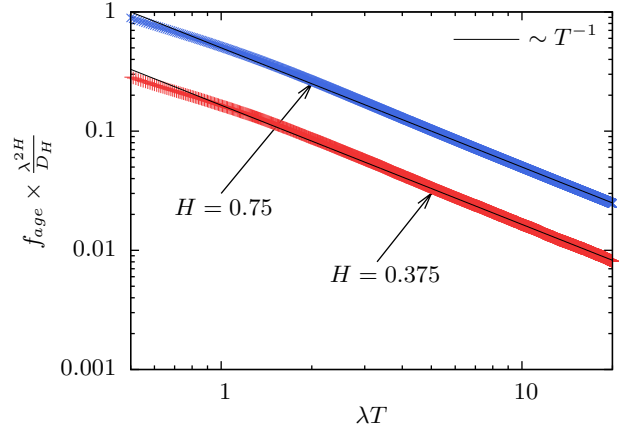}
\caption{Measurement time $T$ dependence for confined FBM, confirming the $1/T$
behavior. Simulation parameters: $h=0.01$ (see Appendix \ref{sec:simulation}),
$\Delta/\tau=0.2$, and $t_a=0$. The results were averaged over $10^5$
trajectories. The simulation for $H=0.75$ started with $x_0^2=25D_H/\lambda^{
2H}$, the one for $H=0.375$ with $x_0^2=9D_H/\lambda^{2H}$.}
\label{fig:fbmkTageing}
\end{figure}

\begin{figure}
\includegraphics{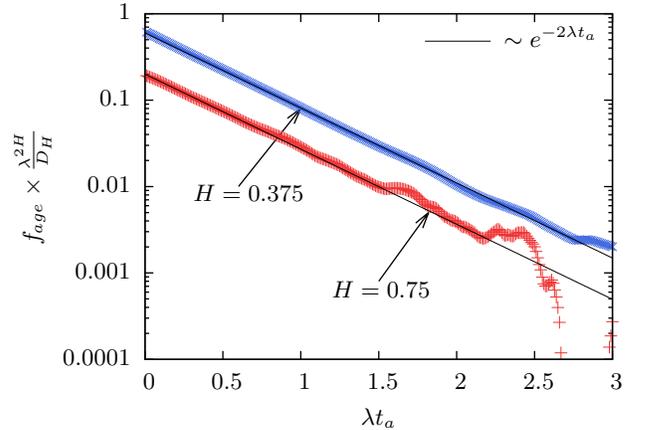}
\caption{Ageing for confined FBM, showing the predicted exponential decay \eqref{eq:fbmk_ageing_ta}.
Simulations used $h=0.01$ (see Appendix \ref{sec:simulation}) and $\Delta/
\tau=0.2$, averaging over $10^5$ trajectories. The simulation for $H=0.75$
started with $x_0^2=25D_H/\lambda^{2H}$, the one for $H=0.375$ with
$x_0^2=9D_H/\lambda^{2H}$.}
\label{fig:agingtafbmk}
\end{figure}

From the explicit result for the time averaged MSD $\langle\overline{\delta^2
(\Delta)}\rangle$ in Eq.~(\ref{fbmklongfunction}) we see that also for this
process we obtain the additive combination (\ref{age_split}) of a stationary
and an ageing contribution. The stationary part is given by
\begin{eqnarray}
\nonumber
f_{\mathrm{st}}&=&2D_H\Delta^{2H}+\frac{2D_H}{\lambda^{2H}}\Gamma(2H+1)\\
\nonumber
&&\hspace*{-1.2cm}-\frac{D_H}{\lambda^{2H}}(\Gamma(2H+1,\lambda\Delta)
e^{\lambda\Delta}+\Gamma(2H+1)e^{-\lambda\Delta})\\
&&\hspace*{-1.2cm}
-\frac{D_H\lambda}{2H+1}\Delta^{2H+1}e^{-\lambda\Delta}M(2H+1;2H+2;\lambda
\Delta),
\label{eq:TAMSDstat}
\end{eqnarray}
where $\Gamma(a,z)$ is the upper incomplete $\Gamma$-function. Interestingly,
as discovered in Ref.~\cite{jeon1} the approach to the stationary value of
the time averaged MSD is of power-law form,
\begin{eqnarray}
\nonumber
f_{\mathrm{st}}&\sim&\frac{2D_H\Gamma(2H+1)}{\lambda^{2H}}-\frac{D_H}{\lambda^{
2H}}\Gamma(2H+1)e^{-\lambda\Delta}\\
&&+\frac{4H(2H-1)D_H}{\lambda^2}\Delta^{2H-2},
\end{eqnarray}
contrasting the exponentially fast relaxation of
the ensemble MSD. Note again the factor of two in the stationary value due to
the definition of $\langle \overline{\delta^2(\Delta)}\rangle$.  We demonstrate
in Figs.~\ref{fig:fbmkTageing} and \ref{fig:agingtafbmk} that the measurement
time dependence of the ageing term is indeed of the form $f_{\mathrm{age}}
\simeq1/T$, and that the ageing time dependence is of the form
\begin{equation}
  f_{\mathrm{age}} \sim x_0^2\exp(-2\lambda t_a). \label{eq:fbmk_ageing_ta}
\end{equation}
Both relations can be shown analytically.

\section{Discussion}
\label{disc}

We studied the FLE and FBM models for anomalous diffusion which are driven
by fractional Gaussian noise. From results for the ensemble averaged first
and second moments, we obtained the associated time averaged MSD $\langle
\overline{\delta^2(\Delta)}\rangle$. The exact expression for $\langle\overline
{\delta^2(\Delta)}\rangle$ was shown to separate into two additive
contributions, a stationary one and an ageing term. The ageing term vanishes at
large measurement times inverse-proportionally. While solutions to regular
Langevin equations with white noise as well as the confined FBM show
exponential relaxation of the time averaged MSD in the ageing time, solutions
to FLEs exhibit power-law
relaxation and power-law ageing with a characteristic exponent between -1 and
-3. Notably, the ageing exponents differ between free and confined FLE motion.

The existence of the finite stationary term of the time averaged MSD
for stochastic processes driven by fractional Gaussian noise contrasts
the properties of the CTRW process, for which the time
averaged MSD decays to zero in the respective limits.  The origin of this
difference is that a CTRW particle effectively grinds to a complete halt after
a sufficiently long time while a free FLE particle keeps spreading. Hence, the
ageing dynamics in CTRWs and the free FLE are well distinguishable. In
particular, it should be noted that free FBM, which is equivalent to overdamped
free FLE motion \cite{deng} does not exhibit ageing at all.

What happens when, instead of fixed initial conditions $v_0$, $x_0$, and $y_0$
as well as ageing times $t_a$ we analyze an ensemble of trajectories with
distributed ageing times or initial conditions? In such cases our ageing
formulas may also be useful. Consider the example of free FLE motion.
Let $\langle\cdot\rangle$ denote the ensemble
average for a fixed ageing time $t_a$ and a fixed initial condition $v_0$. Let
$\langle\cdot\rangle_{\mathrm{dis}}$ denote the ensemble average over
trajectories with distributed ageing times $t_a$ and initial velocities $v_0$,
and let $P$ denote the probability for a given value of $v_0$ and $t_a$. Then
we obtain
\begin{equation}
\left<\overline{\delta^2(\Delta)}\right>_{\mathrm{dis}}=\int\limits_0^\infty
dt_a\int\limits_{-\infty}^\infty dv_0\left<\overline{\delta^2(\Delta;t_a,v_0)}
\right>P(t_a,v_0)
\end{equation}
Hence, the time averaged MSD over an ensemble of trajectories with distributed
initial conditions and ageing times can be calculated by weighting our result
for the time averaged MSD for fixed $t_a$ and $v_0$ with the probability for
that ageing time and initial condition to occur, $P(t_a,v_0)$. This
probability will eventually depend on the experimental setup.

\acknowledgments

Financial support from the Engineering and Physical Sciences Research Council
(EPSRC), the Academy of Finland (FiDiPro scheme), and the CompInt graduate
school at TUM is gratefully acknowledged. RM thanks the Mathematical Institute
of the University of Oxford for financial support as an OCCAM Visiting Fellow.

\begin{appendix}

\section{Simulation Schemes}
\label{sec:simulation}

The simulation schemes for the stochastic differential equations are based
on the method originally proposed by Deng and Barkai \cite{deng} and also
used in Ref.~\cite{jae}. In order to derive the scheme, we first integrate
the dynamic equations \eqref{eq:ffle}, \eqref{eq:flelambda}, and \eqref{eq:fbmk} over time. This leads to
\begin{equation}
\label{volterra}
mv(t)=mv_0-\frac{\overline{\gamma}}{2H-1}\int\limits_0^t(t-\tau)^{2H-1}v(\tau)
d\tau+\eta B_H(t)
\end{equation}
for the free FLE,
\begin{eqnarray}
\nonumber
\lambda\int\limits_0^ty(\tau)d\tau&=&-\overline{\gamma}\int\limits_0^t(t-\tau)^{
2H-2}y(\tau)d\tau\\
&&+\frac{\overline{\gamma}y_0}{2H-1}t^{2H-1}+\eta B_H(t)
\end{eqnarray}
for the overdamped confined FLE, and
\begin{equation}
x(t)=x_0-\lambda \int\limits_0^tx(\tau)d\tau+B_H(t). 
\end{equation}
for confined FBM. These integrals are
interpreted trajectory-wise, in analogy to
Refs.~\cite{mandelbrot,gripenberg,kou}. We then rescale these equations
according to $y\to y/\sqrt{k_B\mathcal{T}/(m\gamma^{1/(2H)})}$ (free FLE motion),
$y\to y/\sqrt{ k_B\mathcal{T}/\lambda}$ (confined FLE motion) and $x\to
x/\sqrt{D_H/\lambda^{2H}}$ (confined FBM). The time is rescaled through $t\to
t/\tau$, where $\tau$ is the intrinsic time scale of each process. FBM is
rescaled by $B_H\to (\tau)^{-H}B_ H$. Dividing the interval $[0,t]$ into $n+1$
intervals of even length $h$, we linearize the fractional derivatives using the
results from Diethelm et al.~\cite{diethelm} 
\begin{equation}
\int\limits_0^{t_{n+1}}(t_{n+1}-\tau)^{\alpha-1}f(\tau)d\tau\approx\frac{h^{
\alpha}}{\alpha(\alpha+1)}\sum_{j=0}^{n+1}a_{j,n+1}f(t_j)
\end{equation}
where
\begin{widetext}
\begin{equation}
a_{j,n+1}=\left\{
\begin{array}{ll}n^{\alpha+1}-(n-\alpha)(n+1)^{\alpha},&j=0\\[0.2cm]
(n-j+2)^{\alpha+1}+(n-j)^{\alpha+1}-2(n-j+1)^{\alpha+1}, &1\le j\le n\\[0.2cm]
1,&j=n+1.\end{array}\right.
\end{equation}
One can then solve for the $v(t_{n+1})$, $y(t_{n+1})$, and $x(t_{n+1})$. This
leads in the case of the free FLE to
\begin{equation}
v(t_{n+1})=\frac{\Gamma(2H+2)}{C}v_0-\frac{h^{2H}}{C}\sum_{j=0}^{n}a_{
j,n+1}v(t_j)+\frac{\Gamma(2H+2)}{C\sqrt{\Gamma(2H+1)D_H}}B_H(t_{n+1})
\end{equation}
with $C = \Gamma(2H+2) + h^{2H}$ and
\begin{equation}
a_{j,n+1}=\left\{\begin{array}{ll}
n^{2H+1}-(n-2H)(n+1)^{2H},&j=0\\[0.2cm]
(n-j+2)^{2H+1}+(n-j)^{2H+1}-2(n-j+1)^{2H+1},&1\le j\le n\\[0.2cm]
1&j=n+1\end{array}\right..
\end{equation}
The velocities obtained by this scheme can then be integrated by the trapezoidal
rule to obtain the particle position. For the overdamped FLE-$\lambda$ we get
\begin{equation}
y(t_{n+1})=-\sum_{j=0}^nd_{n+1,j}y(t_j)+\frac{2y_0}{\frac{1}{H}+\Gamma(2H)
h^{2-2H}}(n+1)^{2H-1}+\frac{2}{\sqrt{D_H\Gamma(2H+1)}\left(h+\frac{2h^{2H-1}}{
\Gamma(2H+1)}\right)}B_H(t_{n+1})
\end{equation}
with
\begin{equation}
d_{n+1,j}=\left\{\begin{array}{ll}
\frac{1}{1+\frac{2h^{2H-2}}{\Gamma(2H+1)}}+\frac{2}{2+\Gamma(2H+1)h^{2-2H}}
\left[n^{2H}-(n+1-2H)(n+1)^{2H-1}\right],&j=0\\[0.6cm]
\frac{2}{1+\frac{2h^{2H-2}}{\Gamma(2H+1)}}+\frac{2}{2+\Gamma(2H+1)h^{2-2H}}
\left[(n-j+2)^{2H}+(n-j)^{2H}-2(n-j+1)^{2H}\right],&1\le j\le n\end{array}
\right.
\end{equation}
and, finally, in the case of confined FBM one obtains
\begin{equation}
x(t_{n+1})=\frac{2-h}{2+h}x(t_n)+\frac{2}{(2+h)\sqrt{D_H}}(B_H(t_{n+1})-
B_H(t_n)).
\end{equation}
\end{widetext}
Hence, we derived expressions for the particle positions or velocities at the
time point $t_{n+1}$ depending on all prior positions or velocities. These
expressions can be used to obtain trajectories of the corresponding stochastic
processes. To simulate FBM, the implementation provided in Ref.~\cite{dieker}
was used. It is based on a method originally developed by Hosking \cite{hosking}.
It should be noted that all the proposed algorithms have time complexity of order
$n^2$.

\section{Solutions to the stochastic differential equations}
\label{app_ex}

To analytically solve equations equations \eqref{eq:ffle},
\eqref{eq:flelambda}, and \eqref{eq:fbmk} we formally derive their solutions in
Laplace space, treating the noise as a well-behaved function of time analogous
to Deng and Barkai \cite{deng}.  Averaging and considering that $\langle \xi(s)
\rangle = 0$ holds for the Laplace transform of fractional Gaussian noise
\eqref{eq:FGN} allows us to identify the first moments
\eqref{eq:first_moment_FFLE}, \eqref{firstfle},
\eqref{eq:OFLE_lambda_first_moment}, and \eqref{eq:first_moment_fbmk}. We then
multiply each formal solution at two different points in Laplace space.  The
average of these products gives us the double Laplace transform of the second
moments.  To obtain explicit expressions, we use the double Laplace transform
of fractional Gaussian noise ($\tau$ and $s$ being the Laplace variables)

\begin{eqnarray}
\langle\xi(s)\xi(\tau)\rangle&=&D_H\Gamma(2H+1)\frac{s^{1-2H}+\tau^{1-2H}}{
s+\tau}.
\end{eqnarray}

The second moments follow as 

\begin{subequations}
\begin{widetext}
\begin{eqnarray}
\nonumber
\langle y(t_1)y(t_2)\rangle&=&\frac{k_B\mathcal{T}}{m}\Big\{t_1^2E_{2H,3}(
-\gamma t_1^{2H})+t_2^2E_{2H,3}(-\gamma t_2^{2H})-(t_1-t_2)^2E_{2H,3}(-\gamma|
t_1-t_2|^{2H})\Big\}\\
&&+\left(v_0^2-\frac{k_B\mathcal{T}}{m}\right)t_1t_2E_{2H,2}(-\gamma t_1^{2H})
E_{2H,2}(-\gamma t_2^{2H})
\label{eq:sec_mom_FFLE}
\end{eqnarray}
for free FLE motion,
\begin{eqnarray}
\nonumber
\langle y(t_1)y(t_2)\rangle&=&\frac{k_B\mathcal{T}}{\lambda}A(|t_2-t_1|)+
\left(y_0^2-\frac{k_B\mathcal{T}}{\lambda}\right)A(t_1)A(t_2)+\left(v_0^2-
\frac{k_B\mathcal{T}}{m}\right)B(t_1)B(t_2)\\
&&+y_0v_0(A(t_1)B(t_2)+A(t_2)B(t_1))
\label{eq:sec_mom_FLElambda}
\end{eqnarray}
for confined FLE motion [compare Eqs.~(\ref{abbs})],
\begin{equation}
\label{eq:sec_mom_ov_FLElambda}
\langle y(t_1)y(t_2)\rangle=\frac{k_B\mathcal{T}}{\lambda}E_{2-2H}(-\frac{
\lambda}{\gamma}|t_2-t_1|^{2-2H})+\left(y_0^2-\frac{k_B\mathcal{T}}{\lambda}
\right)E_{2-2H}(-\frac{\lambda}{\gamma}t_1^{2-2H})E_{2-2H}(-\frac{\lambda}{
\gamma}t_2^{2-2H})
\end{equation}
for overdamped confined FLE motion (note that the definitions of $\gamma$
differ for confined and free FLE motion), and
\begin{eqnarray}
\nonumber
\langle x(t_1)x(t_2)\rangle&=&D_H\lbrace e^{-\lambda t_1}t_2^{2H}+e^{-\lambda t_2}t_1^{2H}
-|t_1-t_2|^{2H}\rbrace+\frac{D_H}{2\lambda ^{2H}}\lbrace e^{-\lambda |t_2-t_1|}\gamma(2H+1,
\lambda t_1)+e^{\lambda |t_2-t_1|}\gamma(2H+1,\lambda t_2) \rbrace\\
\nonumber
&&-\frac{D_H}{2\lambda ^{2H}} e^{\lambda |t_2-t_1|}\gamma(2H+1,\lambda |t_2-t_1|)
-\frac{\lambda D_H}{2(2H+1)}e^{-\lambda (t_2+t_1)} t_1^{2H+1}M(2H+1;2H+2;\lambda t_1)\\
\nonumber
&&-\frac{\lambda D_H}{2(2H+1)}e^{-\lambda (t_2+t_1)}t_2^{2H+1}M(2H+1;2H+2;\lambda t_2) \\
&&+\frac{\lambda D_H}{2(2H+1)}|t_2-t_1|^{2H+1}e^{-\lambda |t_2-t_1|}M(2H+1;2H+2;\lambda |t_2-t_1|)
\label{eq:sec_mom_FBMk}+x_0^2e^{-\lambda (t_1+t_2)},~(t_2>t_1).
\end{eqnarray}
for confined FBM. 
\end{widetext}
\end{subequations}
Taken together, the first moments \eqref{eq:first_moment_FFLE},
\eqref{firstfle}, \eqref{eq:OFLE_lambda_first_moment}, and \eqref{eq:first_moment_fbmk} and the second moments
\eqref{eq:sec_mom_FFLE} to \eqref{eq:sec_mom_FBMk} compose the full solutions
of the stochastic differential equations \eqref{eq:ffle}, \eqref{eq:flelambda},
and \eqref{eq:fbmk} \cite{vankampen}. These solutions provide fundamental insight into the behavior
of the underlying processes. They allow one to interpret the short and long
time correlations, capture the complete dynamics of the equilibration, and
provide the basis to calculate other valuable quantities, such as velocity
correlations or the time averaged MSDs.

\section{Time averaged MSD for confined FBM}

For confined FBM the time averaged MSD takes on the complex form
\begin{widetext}
\begin{eqnarray}
\label{fbmklongfunction}
&&\hspace{-0.5cm}\left<\overline{\delta^2(\Delta)}\right>=2D_H\Delta^{2H}+\frac{D_H}{\lambda ^{2H}}e^{\lambda \Delta}\gamma(2H+1,\lambda \Delta)-\frac{D_H\lambda }{2H+1}\Delta^{2H+1}e^{-\lambda \Delta}M(2H+1;2H+2;\lambda \Delta)\\
\nonumber
&&\hspace{-0.5cm}+\frac{x_0^2}{2\lambda (T-\Delta)}e^{-2\lambda t_a}\big(1-e^{-2\lambda (T-\Delta)}\big)\left\lbrace 1+e^{-\lambda \Delta}\left(e^{-\lambda \Delta}-2\right)\right\rbrace\\
\nonumber
&&\hspace{-0.5cm}+(1-e^{-\lambda \Delta})\frac{D_H}{(T-\Delta)}\Big\{(t_a+T-\Delta)^{2H+1}e^{-\lambda (t_a+T-\Delta)}-t_a^{2H+1}e^{-\lambda t_a}\Big\}\\
\nonumber
&&\hspace{-0.5cm}+(1-e^{\lambda \Delta})\frac{D_H}{(T-\Delta)}\Big\{(t_a+T)^{2H+1}e^{-\lambda (t_a+T)}-(t_a+\Delta)^{2H+1}e^{-\lambda (t_a+\Delta)}\Big\}\\
\nonumber
&&\hspace{-0.5cm}+(1-e^{-\lambda \Delta})\frac{D_H}{2(2H+1)(T-\Delta)}\Big\{(t_a+T-\Delta)^{2H+1}e^{-2\lambda (t_a+T-\Delta)}  M(2H+1;2H+2,\lambda (t_a+T-\Delta)) \\
\nonumber
&&\hspace{5cm}-t_a^{2H+1}e^{-2\lambda t_a}M(2H+1;2H+2,\lambda t_a)\Big\}\\
\nonumber
&&\hspace{-0.5cm}+(1-e^{\lambda \Delta})\frac{D_H}{2(2H+1)(T-\Delta)}\Big\{(t_a+T)^{2H+1}e^{-2\lambda (t_a+T)}M(2H+1;2H+2,\lambda (t_a+T)) \\
\nonumber
&& \hspace{5cm}-(t_a+\Delta)^{2H+1}e^{-2\lambda (t_a+\Delta)}M(2H+1;2H+2,\lambda (t_a+\Delta))\Big\}\\
\nonumber
&&\hspace{-0.5cm}+(1-e^{-\lambda \Delta})\frac{D_H}{\lambda ^{2H}(T-\Delta)}\Bigg\{\gamma(2H+1,\lambda (t_a+T-\Delta))\left(t_a+T-\Delta -\frac{4H-1}{2\lambda }\right)-\gamma(2H+1,\lambda t_a)\left(t_a-\frac{4H-1}{2\lambda }\right)\Bigg\}\\
&&\hspace{-0.5cm}+(1-e^{\lambda \Delta})\frac{D_H}{\lambda ^{2H}(T-\Delta)}\Bigg\{\gamma(2H+1,\lambda (t_a+T))\left(t_a+T-\frac{4H-1}{2\lambda }\right) 
-\gamma(2H+1,\lambda (t_a+\Delta))\left(t_a+\Delta-\frac{4H-1}{2\lambda }\right)\Bigg\}.
\nonumber
\end{eqnarray}
\end{widetext}

\end{appendix}

\end{document}